\documentstyle[11pt,aaspp4,flushrt]{article}

\slugcomment{\it to appear in The Astronomical Journal (April 2000)}
\lefthead{Stern et al.}
\righthead{Radio Properties of $z > 4$ Quasars }

\def\eg{{e.g.,}}
\def\etal{{et al.~}}

\def\deg{\ifmmode {^{\circ}}\else {$^\circ$}\fi}

\def\secper{\ifmmode \rlap.{^{s}}\else $\rlap{.}{^{s}} $\fi}

\def\kms{\ifmmode {\rm\,km\,s^{-1}}\else
    ${\rm\,km\,s^{-1}}$\fi}
\def\kmsMpc{\ifmmode {\rm\,km\,s^{-1}\,Mpc^{-1}}\else
    ${\rm\,km\,s^{-1}\,Mpc^{-1}}$\fi}
\def\ergAcm2{\ifmmode {\rm\,ergs\,cm^{-2}\,{\rm \AA}^{-1}}\else
    ${\rm\,ergs\,cm^{-2}\,\AA^{-1}}$\fi}
\def\ergcm2s{\ifmmode {\rm\,ergs\,cm^{-2}\,s^{-1}}\else
    ${\rm\,ergs\,cm^{-2}\,s^{-1}}$\fi}
\def\ergsHz{\ifmmode {\rm\,ergs\,s^{-1}\,Hz^{-1}}\else
    ${\rm\,ergs\,s^{-1}\,Hz^{-1}}$\fi}
\def\ergs{\ifmmode {\rm\,ergs\,s^{-1}}\else
    ${\rm\,ergs\,s^{-1}}$\fi}
\def\ergsA{\ifmmode {\rm\,ergs\,s^{-1}\,\AA^{-1}}\else
    ${\rm\,ergs\,s^{-1}\,\AA^{-1}}$\fi}
\def\WHz{\ifmmode {\rm\,W\,Hz^{-1}}\else
    ${\rm\,W\,Hz^{-1}}$\fi}
\def\WHzsr{\ifmmode {\rm\,W\,Hz^{-1}\,sr^{-1}}\else
    ${\rm\,W\,Hz^{-1}\,sr^{-1}}$\fi}
\def\ergscmHz{\ifmmode {\rm\,ergs\,cm^{-2}\,Hz^{-1}}\else
    ${\rm\,ergs\,cm^{-2}\,Hz^{-1}}$\fi}

\def\spose#1{\hbox to 0pt{#1\hss}}
\def\simlt{\mathrel{\spose{\lower 3pt\hbox{$\mathchar"218$}}
     \raise 2.0pt\hbox{$\mathchar"13C$}}}
\def\simgt{\mathrel{\spose{\lower 3pt\hbox{$\mathchar"218$}}
     \raise 2.0pt\hbox{$\mathchar"13E$}}}

\def\lya{Ly$\alpha$}

\begin{document}

\title{Radio Properties of $z >4$ Optically-Selected Quasars}

\author{Daniel Stern\altaffilmark{1}, S.~G.~Djorgovski\altaffilmark{2},
R.~A.~Perley\altaffilmark{3}, \\ Reinaldo
R.~de~Carvalho\altaffilmark{4}, \& J.~V.~Wall\altaffilmark{5}}

\altaffiltext{1}{Department of Astronomy, University of California,
Berkeley, CA 94720; {\tt dan@slothrop.berkeley.edu}}

\altaffiltext{2}{Palomar Observatory, Caltech, Pasadena, CA 91125; {\tt
george@deimos.caltech.edu}}

\altaffiltext{3}{NRAO, P.O. Box 0, Socorro, NM 87801-0387; {\tt
rperley@nrao.edu}}

\altaffiltext{4}{Observatorio Nacional, Rio de Janeiro, Brazil; {\tt
reinaldo@voyager.on.br}}

\altaffiltext{5}{Department of Astrophysics, University of Oxford,
Keble Road, Oxford, UK OX13RH; {\tt jvw@astro.ox.ac.uk}}

\begin{abstract}

We report on two programs to address differential evolution between the
radio-loud and radio-quiet quasar populations at high ($z > 4$)
redshift.  Both programs entail studying the radio properties of
optically-selected quasars.  First, we have observed 32
optically-selected, high-redshift ($z > 4$) quasars with the VLA at
6~cm (5~GHz).  These sources comprise a statistically complete and
well-understood sample.   We detect four quasars above our 3$\sigma$
limit of $\approx$ 0.15~mJy, which is sufficiently sensitive to detect
all radio-loud quasars at the probed redshift range.  Second, we have
correlated 134 $z > 4$ quasars, comprising all such sources that we are
aware of as of mid-1999, with FIRST and NVSS.  These two recent 1.4~GHz
VLA sky surveys reach 3$\sigma$ limits of approximately 0.6~mJy and
1.4~mJy respectively.  We identify a total of 15 $z > 4$ quasars, of
which six were not previously known to be radio-loud.  The depth of
these surveys does not reach the radio-loud/radio-quiet demarcation
luminosity density ($L_{\rm 1.4~GHz} = 10^{32.5}~ h_{50}^{-2} \ergsHz$)
at the redshift range considered; this correlation therefore only
provides a lower limit to the radio-loud fraction of quasars at
high-redshift.  The two programs together identify eight new radio-loud
quasars at $z > 4$, a significant increase over the seven currently in
the published literature.  We find no evidence for radio-loud fraction
depending on optical luminosity for $-25 > M_B > -28$ at $z \simeq 2$,
or for $-26 > M_B > -28$ at $z > 4$.  Our results also show no
evolution in the radio-loud fraction between $z \simeq 2$ and $z > 4$
($-26 > M_B > -28$).

\end{abstract}


\section{Introduction}

Already in the late 1960s it was becoming apparent that the rapid
increase in quasar comoving space densities with redshift did not
continue beyond $z \sim 2$.  Several authors suggested the existence of
a redshift cutoff beyond which a real decrease in quasar densities
occurs, unrelated to observational selection effects
\markcite{Sandage:72}(\eg Sandage 1972).  Numerous studies have now shown that this
is indeed the case.  For example, \markcite{Schmidt:95b}Schmidt, Schneider, \&  Gunn (1995a), applying the
$V/V_{\rm max}$ test to a well-defined sample of 90 quasars detected by
their \lya\ emission in the Palomar Transit Grism Survey, conclude that
the comoving quasar density decreases from $z = 2.75$ to $z = 4.75$.
\markcite{Warren:94}Warren, Hewett, \& Osmer (1994) show that the quasar space density falls by a factor
of a few between $z = 3.3$ and $z = 4.0$.  \markcite{Shaver:96b}Shaver {et~al.} (1996b) show a
similar decrease in space density exists for radio-loud quasars,
implying that the decline is not simply due to obscuration by
intervening galaxies.


An unresolved question, however, is how the decline compares
between the optically-selected and radio-selected populations.  Two
definitions are generally used to demarcate radio-quiet and radio-loud
quasars.  One criterion is the radio-optical ratio $R_{\rm r-o}$ of the
specific fluxes at restframe 6~cm (5~GHz) and 4400
\AA\ \markcite{Kellerman:89}(Kellerman {et~al.} 1989); radio-loud sources typically have $R_{\rm
r-o}$ in the range $10 - 1000$ while most radio-quiet sources have $0.1
< R_{\rm r-o} < 1$.  However, \markcite{Peacock:86}Peacock, Miller, \& Longair (1986) and \markcite{Miller:90}Miller, Peacock, \& Mead (1990)
point out that $R_{\rm r-o}$ is physically meaningful as a
discriminating parameter {\em only} if radio and optical luminosities
are linearly correlated.  This is apparently not the case;  no linear
correlation has been observed for radio-loud quasars \markcite{Stocke:92}(Stocke {et~al.} 1992)
and the fraction of optically-faint, optically-selected quasars which
are radio-loud would be higher if radio and optical properties were
linearly correlated \markcite{Peacock:86, Miller:90}(Peacock {et~al.} 1986; Miller {et~al.} 1990).

The second definition, which can result in ambiguous taxonomy for some
quasars, is to divide the populations at some restframe radio
luminosity \markcite{Miller:90, Schneider:92}(Miller {et~al.} 1990; Schneider {et~al.} 1992).  Different authors use
slightly different luminosities to discriminate the populations; for
what follows, we identify radio-loudness using the \markcite{Gregg:96}Gregg {et~al.} (1996)
cutoff value for the 1.4~GHz specific luminosity, $L_{\rm 1.4 GHz} =
10^{32.5}~ h_{50}^{-2}$ \ergsHz ($\approx 10^{24}~ h_{50}^{-2}$
\WHzsr).

Differential evolution between the radio-quiet and radio-loud quasar
populations would be a fascinating result, which could change (or
challenge) our understanding of different types of active galactic nuclei
(AGN).  For example, \markcite{Wilson:95}Wilson \& Colbert (1995) propose a model whereby radio-loud
AGN are products of coalescing supermassive black holes in galaxy mergers;
such a process would result in rapidly spinning black holes, capable of
generating highly-collimated jets and powerful radio sources.  Such a
model could naturally explain a lag (if there is indeed one) between the
appearance of powerful radio-loud quasars in the early Universe and the
more common radio-quiet quasars.

Several groups have reported on followup of optically-selected quasar
surveys at radio frequencies \markcite{Kellerman:89, Miller:90,
Schneider:92, Visnovsky:92, Hooper:95, Schmidt:95, Goldschmidt:99}(\eg Kellerman {et~al.} 1989; Miller {et~al.} 1990; Schneider {et~al.} 1992; Visnovsky {et~al.} 1992; Hooper {et~al.} 1995; Schmidt {et~al.} 1995b; Goldschmidt {et~al.} 1999).
Typically between 10\%\ and 40\%\ of the quasars are detected in the
radio.  One notable exception is \markcite{Kellerman:89}Kellerman {et~al.} (1989) who, using the
Very Large Array (VLA), detect more than 80\%\ of 114 quasars
observed from the Palomar-Green Bright Quasar Survey
\markcite{Schmidt:83}(BQS; Schmidt \& Green 1983), a relatively low-redshift sample.
\markcite{Visnovsky:92}Visnovsky {et~al.} (1992) report on radio observations of 124 quasars in
the redshift range $1 < z < 3$ selected from the Large Bright Quasar
Survey \markcite{Hewett:95}(LBQS; Hewett, Foltz, \& Chaffee 1995).  The sample, chosen to complement
the predominantly low-redshift BQS, leads them to conclude that the
fraction of radio-loud quasars decreases with increasing redshift.  A
similar conclusion is reached by \markcite{Schneider:92}Schneider {et~al.} (1992), who
report on 5~GHz VLA observations of 22 optically-selected quasars at
$z > 3.1$.  \markcite{LaFranca:94}{La~Franca} {et~al.} (1994), combining the results of several radio
followup studies of optically-selected quasars, also concludes that the
radio-loud fraction of optically-selected quasars decreases with
increasing redshift; however, this conclusion is statistically robust
{\em only} when the BQS sample is included.

\markcite{Hooper:95}Hooper {et~al.} (1995) report on 8.4~GHz VLA observations of an additional 132
LBQS quasars, supplementing the sample discussed by \markcite{Visnovsky:92}Visnovsky {et~al.} (1992).
Contrary to the results discussed above, \markcite{Hooper:95}Hooper {et~al.} (1995) find that the
radio-loud quasar fraction of the LBQS exhibits a peak around $z \approx
1$, is constant at $\approx$ 10\%\ until $z \approx 2.5$, and then {\em
increases} substantially to $z = 3.4$, the highest redshift in the LBQS.
They suggest the enhanced radio-loud fraction may be the result of
increased radio-loud fraction at bright optical absolute magnitude
($M_B \simlt -27.4$).

The most recent thorough analysis of the issue is presented by
\markcite{Goldschmidt:99}Goldschmidt {et~al.} (1999), who report on 5~GHz VLA observations of 87
optically-selected quasars from the Edinburgh Quasar Survey.  They
combine these results with all published radio surveys of
optically-selected quasar samples other than the BQS sample which has
been shown to be incomplete by a factor of three
\markcite{Goldschmidt:92}(Goldschmidt {et~al.} 1992).  \markcite{Miller:93}Miller, Rawlings, \& Saunders (1993) suggest that this
incompleteness is not random with respect to radio properties, implying
that it is inappropriate to use the BQS for determinations of the
radio-loud fraction.  \markcite{Goldschmidt:99}Goldschmidt {et~al.} (1999) further fortify their
sample by correlating several large quasar surveys with two recent,
large-area, deep radio surveys:  Faint Images of the Radio Sky at
Twenty centimeters \markcite{Becker:95}(FIRST; Becker, White, \& Helfand 1995) and the NRAO VLA Sky
Survey \markcite{Condon:98}(NVSS; Condon {et~al.} 1998).  This work samples the optical
luminosity--redshift plane more thoroughly than any previous analysis.
Considering only quasars in the absolute magnitude interval $-25.5 \geq
M_B \geq -26.5$, they find hints that the radio-loud fraction decreases
with increasing redshift, though they note that the result is {\em not}
statistically significant in the narrow $M_B$ range considered.

In short, the existence of differential evolution between the
radio-quiet and radio-loud quasar populations remains uncertain.  Some
researchers find evidence that the radio-loud fraction of
optically-selected quasars decreases with increasing redshift
\markcite{Visnovsky:92}(\eg Visnovsky {et~al.} 1992), while others find no (or only marginal)
evidence for evolution \markcite{LaFranca:94, Goldschmidt:99}(\eg {La~Franca} {et~al.} 1994; Goldschmidt {et~al.} 1999), and
yet other researches find that the radio-loud fraction {\em increases}
with increasing redshift \markcite{Hooper:95}(\eg Hooper {et~al.} 1995).  We attempt to
re-address the issue by providing a more accurate measure of the
radio-loud fraction at $z > 4$.  In \S\ref{sec_qsoz4_obs} we report on
5~GHz VLA observations of a sample of 32 $z > 4$ optically-selected
quasars.  The \markcite{Schneider:92}Schneider {et~al.} (1992) 5~GHz VLA study includes 13 $z > 4$
quasars; we include these sources in our analysis.  In
\S\ref{sec_qsoz4_opt} we report on correlations of larger, more recent
surveys of $z > 4$ quasars with deep, large-area radio surveys.  
\S\ref{sec_qsoz4_result} sets up the methodology for studying
the radio-loud fraction, discussed in \S\ref{sec_qsoz4_discuss}.

Throughout we adopt the cosmology consistent with previous work in this
field:  an Einstein-de~Sitter cosmology with $H_0 = 50~ h_{50}$
\kmsMpc, $\Omega = 1$, and vanishing cosmological constant, $\Lambda =
0$.

\begin{table}[t!]
\caption{Optical and Radio Properties of the High-Redshift VLA Sample}
\footnotesize
\begin{center}
\begin{tabular}{lcccccc}
\hline\hline
&
&
$r$ &
R.A. &
Dec. &
$M_B$ &
$S_{\rm 5 GHz}$ \\
Quasar &
$z$ &
(mag) &
(J2000) &
(J2000) &
(mag) &
(mJy) \\
\hline
PSS0003+2730 &  4.26 &  19.0 & 00 03 23.10 &   +27 30 23.0 & $-27.54$ & $<0.171$ \\
PSS0030+1702 &  4.28 &  19.3 & 00 30 16.39 &   +17 02 40.2 & $-27.27$ & $<0.141$ \\ 
PSS0034+1639 &  4.38 &  19.5 & 00 34 54.82 &   +16 39 19.5 & $-27.23$ & $<0.171$ \\
PSS0059+0003 &  4.16 &  19.5 & 00 59 22.80 &   +00 03 01.2 & $-26.90$ & $<0.138$ \\
PSS0106+2601 &  4.32 &  19.4 & 01 06 00.80 &   +26 01 02.5 & $-27.23$ & $<0.144$ \\
BRI0103+0032 &  4.43 &  18.6 & 01 06 19.14 &   +00 48 23.1 & $-28.22$ & $<0.216$ \\
PSS0117+1552 &  4.24 &  18.6 & 01 17 31.17 &   +15 52 16.3 & $-27.91$ & $<0.183$ \\
PSS0132+1341 &  4.16 &  19.3 & 01 32 09.85 &   +13 41 39.5 & $-27.10$ & $<0.165$ \\
PSS0134+3307 &  4.52 &  18.8 & 01 34 21.60 &   +33 07 56.5 & $-28.18$ & $<0.165$ \\
PSS0137+2837 &  4.30 &  19.0 & 01 37 12.30 &   +28 37 34.5 & $-27.60$ & $<0.171$ \\
PSS0152+0735 &  4.07 &  19.6 & 01 52 11.10 &   +07 35 50.2 & $-26.70$ & $<0.165$ \\
BRI0151$-$0025 &4.20 &  18.9 & 01 53 39.61 & $-$00 11 04.9 & $-27.55$ & $7.96 \pm 0.110$ \\
BRI0241$-$0146 &4.04 &  18.2 & 02 44 01.85 & $-$01 34 03.0 & $-28.07$ & $<0.162$ \\
PSS0244$-$0108 &4.00 &  19.0 & 02 44 57.18 & $-$01 08 08.6 & $-27.25$ & $<0.180$ \\
PSS0248+1802 &  4.43 &  18.4 & 02 48 54.28 &   +18 02 49.3 & $-28.42$ & $0.39 \pm 0.052$ \\
PSS0747+4434 &  4.42 &  18.1 & 07 47 50.00 &   +44 34 16.0 & $-28.71$ & $<0.126$ \\
BRI0952$-$0115 &4.43 &  18.7 & 09 55 00.08 & $-$01 30 06.9 & $-28.12$ & $<0.120$ \\
BRI1013+0035 &  4.38 &  18.8 & 10 15 48.96 &   +00 20 19.0 & $-27.93$ & $<0.114$ \\
PSS1048+4407 &  4.45 &  19.2 & 10 48 46.62 &   +44 07 12.7 & $-27.66$ & $<0.159$ \\
BRI1050$-$0000 &4.29 &  18.8 & 10 53 20.43 & $-$00 16 49.6 & $-27.78$ & $10.2 \pm 0.044$ \\
PSS1057+4555 &  4.12 &  17.7 & 10 57 56.39 &   +45 55 51.9 & $-28.65$ & $<0.102$ \\
PSS1159+1337 &  4.08 &  18.5 & 11 59 06.48 &   +13 37 37.7 & $-27.81$ & $<0.123$ \\
PSS1317+3531 &  4.36 &  19.1 & 13 17 43.21 &   +35 31 31.1 & $-27.60$ & $<0.132$ \\
PSS1430+2828 &  4.30 &  19.3 & 14 30 31.63 &   +28 28 34.3 & $-27.30$ & $<0.108$ \\
PSS1435+3057 &  4.35 &  19.3 & 14 35 23.52 &   +30 57 22.5 & $-27.38$ & $<0.108$ \\
PSS1438+2538 &  4.25 &  19.5 & 14 38 35.58 &   +25 38 32.1 & $-27.02$ & $<0.105$ \\
PSS1443+2724 &  4.42 &  19.3 & 14 43 31.17 &   +27 24 37.0 & $-27.51$ & $<0.114$ \\
PSS1456+2007 &  4.25 &  19.5 & 14 56 28.97 &   +20 07 27.1 & $-27.02$ & $1.30 \pm 0.049$ \\
PSS1618+4125 &  4.21 &  19.6 & 16 18 22.73 &   +41 25 59.7 & $-26.86$ & $<0.129$ \\
PSS1646+5514 &  4.05 &  18.1 & 16 46 56.29 &   +55 14 46.7 & $-28.18$ & $<0.105$ \\
PSS1721+3256 &  4.03 &  19.2 & 17 21 06.74 &   +32 56 35.7 & $-27.06$ & $<0.138$ \\
PSS2122$-$0014 &4.18 &  19.1 & 21 22 07.50 & $-$00 14 45.0 & $-27.32$ & $<0.243$ \\
\hline
\end{tabular}
\end{center}
\medskip

\emph{Notes.---}  Radio positions are listed for detected sources; all
other positions are from the optical.  Where no radio source was
detected within 10\arcsec\ of the quasar optical position, we quote the
3$\sigma$ noise measured at the center of the VLA map as an upper limit
to the 5~GHz flux density.  See text for description of $M_B$
determination.  PSS quasars are from the Palomar multicolor survey.
BRI quasars are from the APM survey.  Optical coordinates for
BRI1050$-$0000 are from \markcite{StorrieLombardi:96}Storrie-Lombardi
\etal (1996), which are $\approx$ 50\arcsec\ offset from the
coordinates reported by \markcite{Smith:94}Smith \etal (1994).

\label{qsoz4tab1}
\end{table}
\normalsize

\section{Observations}
\label{sec_qsoz4_obs}

\subsection{VLA Sample Selection}

The 32 $z > 4$ quasars which comprise our VLA sample are listed in
Table~\ref{qsoz4tab1}.  The sample is from the Palomar multicolor
survey \markcite{Kennefick:95, Djorgovski:99}(Kennefick, Djorgovski,
\&  de~Calvalho 1995; Djorgovski {et~al.} 1999) and represents all $z >
4$ quasars found from that survey at the time of the radio
observations.  Six of the quasars are equatorial objects found
previously in a similar search by the Automated Plate Machine (APM)
group \markcite{Irwin:91}(Irwin, McMahon, \& Hazard 1991) which fall
within the Palomar multicolor survey selection criteria.  These 32
sources comprise a statistically complete and well-understood sample of
optically-selected $z > 4$ quasars;  they should be an unbiased set as
far as the radio properties are concerned.

Calculation of the absolute $B$ magnitudes listed in Table~\ref{qsoz4tab1}
is discussed in \S\ref{sec_qsoz4_result}.

\subsection{VLA Observations}

The VLA observations were made on UT 1997 March 14, from 01:30 to 17:30
LST.  The targets were observed at 4835 and 4885~MHz.  Each frequency
was observed with 50~MHz bandwidth and both polarizations.  The
configuration was `B', with a maximum baseline of approximately 10~km
and a resolution of $\approx 1\farcs3$.  Observations were made in
nodding mode, in which we oscillated between 40~$s$ observations of a
calibrator and 240~$s$ observations on a target.  This procedure allowed
us to better track and remove atmospheric fluctuations.  Calibration
and imaging followed standard procedures.  Each target was observed
for $30 - 45$~min, including calibration.  The rms noise was typically
between 30 and 60 $\mu$Jy.

We detect four quasars from this sample in the VLA imaging, where a
match is liberally defined to correspond to a radio source lying within
10\arcsec\ of the optical source.  In fact, the four matches all have
optical-to-radio positional differences less than 2\arcsec, implying
that the identifications are unlikely to be spurious.  Three sources
represent newly identified high-redshift, radio-loud quasars.  The
quasar BRI1050$-$0000 was previously identified as a $10.6 \pm 0.2$~mJy
source at 4.8~GHz by \markcite{McMahon:94}McMahon {et~al.} (1994).  The source PSS1618+4125 has a
4$\sigma$ radio detection 16\farcs0 away from the optical
identification; we do not consider this a positive match.

\section{Optically-Selected $z > 4$ Quasars in FIRST and NVSS}
\label{sec_qsoz4_opt}

To augment the targeted VLA sample discussed above, we have correlated
an updated list of all 134 $z > 4$ quasars known to us in mid-1999 with
two recent, large-area radio surveys.  The FIRST survey
\markcite{Becker:95}(Becker {et~al.} 1995), which is still in
production mode, is a radio map of the northern celestial sphere at
21~cm (1.4~GHz), reaching a typical limiting flux density of 1.0~mJy
(5$\sigma$).  The 1.4~GHz NVSS survey \markcite{Condon:98}(Condon
{et~al.} 1998) covers a larger area to a shallower flux density limit;
the 5$\sigma$ limit of NVSS is 2.25~mJy.  For luminosity distance $d_L$
and radio spectral index $\alpha$ ($S_\nu \propto \nu^\alpha$), the
restframe specific luminosity $L_\nu = 4 \pi d_L^2 S_\nu / (1 + z)^{1 +
\alpha}$, where both $L_\nu$ and $S_\nu$ are measured at the same
frequency.  At $z = 4$ and for a typical quasar spectral index of
$\alpha = -0.5$, the FIRST survey reaches a 3$\sigma$ limiting 1.4~GHz
specific luminosity of $\log L_{1.4}~ (h_{50}^{-2} \ergsHz) = 32.6$.
The comparable limit for the NVSS survey is $\log L_{1.4}~ (h_{50}^{-2}
\ergsHz) = 32.9$.  Therefore, using the radio luminosity definition of
radio loudness, these surveys incompletely sample radio-loud quasars at
high redshift.


\begin{table}[t!]
\caption{Optically-Selected, High-Redshift Quasars Detected by FIRST}
\footnotesize
\begin{center}
\begin{tabular}{lcccccccc}
\hline\hline
&
&
$r$ &
R.A. &
Dec. &
$M_B$ &
$S_{\rm 1.4 GHz}$ &
Offset &
\\
Quasar &
$z$ &
(mag) &
(J2000) &
(J2000) &
(mag) &
(mJy) &
(arcsec) &
Ref \\
\hline
BRI0151$-$0025&4.20 & 18.9 & 01 53 39.616 &$-$00 11 04.63& $-27.55$ & $4.75 \pm 0.15$ & 0.90 & 1$-\star$ \\
PSS1057+4555 & 4.12 & 17.7 & 10 57 56.324 & +45 55 53.04 & $-28.65$ & $1.38 \pm 0.13$ & 2.13 & 2$-\star$ \\
FIRST~J141045+34009 & 4.35 & 19.6 & 14 10 45.758 & +34 09 09.68 & $-27.10$ & $2.07 \pm 0.15$ & 0.84 & 3 \\
GB1428+4217  & 4.72 & 20.9 & 14 30 23.740 & +42 04 36.52 & $-26.44$ & $215.62 \pm 0.15$&0.41 & 4 \\
FIRST~J145224+335424 & 4.12 & 20.4 & 14 52 24.254 & +33 54 24.64 & $-25.95$ & $6.95 \pm 0.13$ & 3.02 & 3 \\
FIRST~J171356+421808 & 4.23 & 19.0 & 17 13 56.150 & +42 18 08.64 & $-27.49$ & $2.80 \pm 0.15$ & 0.00 & 3 \\
\hline
\end{tabular}
\end{center}
\medskip

\emph{Notes.---}  Newly identified high-redshift, radio-loud quasars
are indicated with a $\star$.  Positions are from the FIRST catalog.
References:  (1) \markcite{Smith:94}Smith \etal (1994);  (2)
\markcite{Kennefick:95}Kennefick \etal (1995); (3)
\markcite{Stern:99c}Stern \etal (1999); (4) \markcite{Hook:98}Hook \&
McMahon (1998).

\label{qsoz4tab2}
\end{table}
\normalsize

Of the 134 $z > 4$ quasars, 51 reside in portions of the celestial
sphere observed thus far by FIRST.  We deemed a radio detection in
FIRST to be associated with a high-redshift quasar if it lay within
10\arcsec\ of the quasar optical position.  Six high-redshift,
radio-loud quasars were identified, of which four were previously known
(see Table~\ref{qsoz4tab2}).  One, GB1428+4217 \markcite{Hook:98}(Hook
\& McMahon 1998), is the highest-redshift radio-loud quasar known
currently ($z = 4.72$); it was initially identified on the basis of
being a strong, flat-spectrum radio source coincident with an
unresolved, red, stellar, optical source.  The three sources with VLA
nomenclature were initially identified by correlating the FIRST survey
with the digitized 2{\it nd} generation Palomar sky survey
\markcite{Stern:99c}(Stern {et~al.} 1999).  Two new, high-redshift,
radio-loud quasars have been identified:  BRI0151$-$0025 at $z=4.20$
and PSS1057+3409 at $z=4.12$.  The former was also identified in the
targeted 5~GHz survey discussed above (\S\ref{sec_qsoz4_obs}), implying
a radio spectral index $\alpha = 0.41$.  The latter was {\em
undetected} in the targeted 5~GHz survey, implying an unusually steep
spectral index, $\alpha < -2.05$ (3 $\sigma$) or significant
variability on the time scale corresponding to the epochs of the two
radio surveys.


\begin{table}[t!]
\caption{Optically-Selected, High-Redshift Quasars Detected by NVSS}
\footnotesize
\begin{center}
\begin{tabular}{lcccccccc}
\hline\hline
&
&
$r$ &
R.A. &
Dec. &
$M_B$ &
$S_{\rm 1.4 GHz}$ &
Offset &
\\
Quasar &
$z$ &
(mag) &
(J2000) &
(J2000) &
(mag) &
(mJy) &
(arcsec) &
Ref \\
\hline
PC0027+0525  & 4.10 & 21.5 & 00 29 50.18  & +05 42 12.2  & $-24.82$ & $4.8 \pm 0.5$ & 8.47 & 1$-\star$ \\
PSS0121+0347 & 4.13 & 17.9 & 01 21 26.22  & +03 47 06.3  & $-28.46$ & $78.6 \pm 2.4$& 0.76 & 2$-\star$ \\
BRI0151$-$0025&4.20 & 18.9 & 01 53 39.71  &$-$00 11 12.1 & $-27.55$ & $3.0 \pm 0.5$ & 7.57 & 3$-\star$ \\
RXJ1028.6$-$0844&4.28&18.9 & 10 28 38.82  &$-$08 44 38.6 & $-27.67$ & $269.8 \pm 8.1$ & 22.4 & 4 \\
BRI1050$-$0000&4.29 & 18.8 & 10 53 20.28  &$-$00 16 51.0 & $-27.78$ & $9.7 \pm 0.6$ & 3.04 & 5,6 \\
BRI1302$-$1404&4.04 & 18.6 & 13 05 25.16  &$-$14 20 41.0 & $-27.67$ & $20.6 \pm 0.8$  & 0.61 & 7$-\star$ \\
FIRST~J141045+340909 & 4.36 & 19.6 & 14 10 45.87  & +34 09 22.4  & $-27.10$ & $3.5 \pm 0.15$ & 13.5 & 8 \\
GB1428+4217  & 4.72 & 20.9 & 14 30 23.69  & +42 04 36.8  & $-26.44$ & $211.1 \pm 6.3$ & 1.10 & 9 \\
FIRST~J145224+335424 & 4.12 & 20.4 & 14 52 24.11  & +33 54 23.0  & $-25.95$ & $8.6 \pm 0.5$ & 2.42 & 8 \\
GB1508+5714  & 4.30 & 18.9 & 15 10 02.90  & +57 02 43.3  & $-27.70$ & $202.4 \pm 6.1$ & 22.3 & 10 \\
FIRST~J171356+421808 & 4.23 & 19.0 & 17 13 56.37  & +42 18 06.3  & $-27.49$ & $2.9 \pm 0.5$ & 3.40 & 8 \\
GB1713+2148  & 4.01 & 21.0 & 17 15 21.16  & +21 45 34.2  & $-25.25$ & $447.2 \pm 15.0$ & 2.02 & 9 \\
RXJ1759.4+6638&4.32 & 19.0 & 17 59 26.82  & +66 38 43.5  & $-27.63$ & $3.6 \pm 0.7$ & 11.6 & 11$-\star$ \\
PC2331+0216  & 4.10 & 20.0 & 23 34 32.1   & +02 33 16.7  & $-26.32$ & $2.7 \pm 0.5$ & 6.51 & 12,13 \\
\hline
\end{tabular}
\end{center}
\medskip

\emph{Notes.---}  Newly identified high-redshift, radio-loud quasars
are indicated with a $\star$.  Positions are from the NVSS catalog.
References:  (1) \markcite{Schneider:97}Schneider \etal (1997); (2)
\markcite{Djorgovski:99}Djorgovski \etal (1999); (3)
\markcite{Smith:94}Smith \etal (1994); (4)
\markcite{Zickgraf:97}Zickgraf \etal (1997); (5)
\markcite{Smith:94}Smith \etal (1994); (6) \markcite{McMahon:94}McMahon
\etal (1994); (7) \markcite{StorrieLombardi:96}Storrie-Lombardi \etal
(1996); (8) \markcite{Stern:99c}Stern \etal (1999); (9)
\markcite{Hook:98}Hook \& McMahon (1998); (10) \markcite{Hook:95}Hook
\etal (1995); (11) \markcite{Henry:94}Henry \etal (1994); (12)
\markcite{Schneider:89}Schneider, Schmidt, \& Gunn (1989); (13)
\markcite{Schneider:92}Schneider \etal (1992).  Optical coordinates for
BRI1050$-$0000 are from \markcite{StorrieLombardi:96}Storrie-Lombardi
\etal (1996), which are $\approx$ 50\arcsec\ offset from the
coordinates reported by \markcite{Smith:94}Smith \etal (1994).

\label{qsoz4tab3}
\end{table}
\normalsize

We have also correlated the list of $z > 4$ quasars with the NVSS
catalog.  129 of the 134 quasars are in portions of the sky surveyed by
NVSS.  A radio detection in NVSS was deemed associated with a
high-redshift quasar if it lay within 30\arcsec\ of the quasar optical
position.  We found a total of 14 radio identifications (see Table~3);
6 of these quasars were initially discovered on the basis of their
radio emission \markcite{Hook:95, Hook:98, Stern:99c}(Hook {et~al.}
1995; Hook \& McMahon 1998; Stern {et~al.} 1999) and three were
previously known to be radio-loud \markcite{Schneider:92, McMahon:94,
Zickgraf:97}(Schneider {et~al.} 1992; McMahon {et~al.} 1994; Zickgraf
{et~al.} 1997).  Five new high-redshift, radio-loud quasars have been
identified, of which two were identified in the targeted 5~GHz survey
(\S\ref{sec_qsoz4_obs}):  BRI0150$-$0025 ($\alpha = 0.77$) and
BRI1050$-$0000 ($\alpha = 0.04$).

Although 30\arcsec\ is a rather large matching radius, the surface
density of sources in NVSS is $\sim 60$ deg$^{-2}$, so that the
probability of a chance coincidence of an NVSS source with an optical
source is only 0.013.  We therefore expect that 1.7 of our radio
identifications are spurious.  We note that of the four radio--optical
offsets larger than 10\arcsec, the three largest correspond to known
radio-loud quasars, illustrating that the 30\arcsec\ search radius was
appropriate.  The fourth largest radio--optical offset belongs to
RXJ1759.4+6638, an X-ray selected quasar identified from deep {\it
ROSAT} observations near the North Ecliptic Pole \markcite{Henry:94}(Henry {et~al.} 1994).
NVSS reports a 3.6~mJy source 11\farcs6 away.  Sensitive VLA 1.5~GHz
C-array mapping of this region by \markcite{Kollgaard:94}Kollgaard {et~al.} (1994) finds no object
brighter than 0.5~mJy associated with the source, apparently at
odds with the NVSS results.

We are aware of only one $z > 4$ radio-loud quasar not included by
these exercises:  PKS1251$-$407 at $z = 4.46$ \markcite{Shaver:96}(Shaver, Wall, \&  Kellerman 1996a) which
resides too far south to be within the VLA surveys. 

The list of $z > 4$ quasars comes from diverse sources, limiting the
significance of these detections to studies of the quasar luminosity
function at high redshift.  However, 107 of the 129 $z > 4$ quasars
were initially identified from optical surveys, either the Palomar
multicolor survey \markcite{Djorgovski:99}(Djorgovski {et~al.} 1999), the APM survey
\markcite{Irwin:91}(Irwin {et~al.} 1991), or the long-term program of \markcite{Schneider:89,
Schneider:97}Schneider {et~al.} (1989, 1997).  These surveys should be largely unbiased with respect
to radio properties.  In the following analysis, FIRST/NVSS
detections of these sources will provide lower limits to the radio-loud
quasar fraction at high-redshift.

\section{Results}
\label{sec_qsoz4_result}

\subsection{1.4~GHz Specific Luminosity}

\begin{figure}[!t]\begin{center}
\plotfiddle{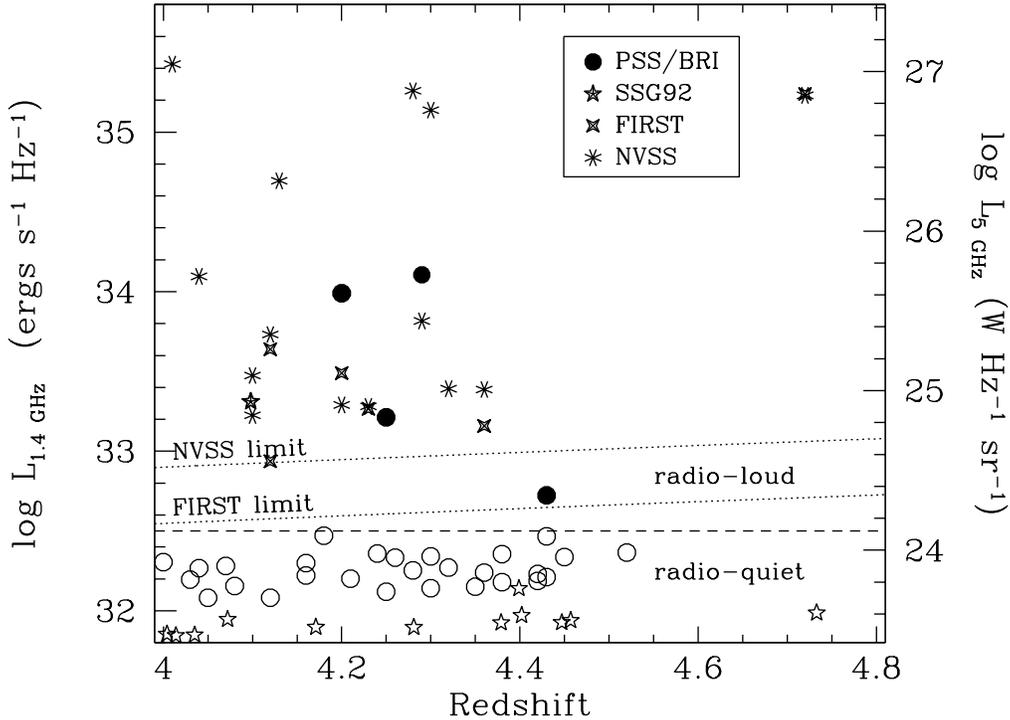}{3.2in}{-90}{50}{50}{-200}{285}
\end{center}

\caption[$L_{\rm 1.4~GHz}$ plotted against $z$ for $z > 4$ quasars]
{$L_{\rm 1.4~GHz}$ plotted against $z$ for $z > 4$ quasars.  Legend
relates symbol to radio source catalog.  3$\sigma$ limits of NVSS and
FIRST surveys are illustrated with dotted lines.  PSS/BRI refers to the
5~GHz deep imaging discussed in \S\ref{sec_qsoz4_obs}.  SSG92 refers to
the \markcite{Schneider:92}Schneider {et~al.} (1992) sample.
Non-detections are indicated with open symbols.  Horizontal dashed line
indicates adopted dividing line between radio-loud quasars and
radio-quiet quasars.  The k-corrections have been calculated assuming
$\alpha = - 0.5$.}

\label{fig_qsoz4_Lz}
\end{figure}

The specific luminosity at 1.4~GHz is straight-forward to calculate.
For $S_\nu \propto \nu^\alpha$ at radio frequencies, \begin{equation}
L_\nu (\nu_1) = 4 \pi d_L^2 { {1} \over {(1 +z)^{1 + \alpha}}} \left(
{{\nu_1} \over {\nu_2}} \right) ^\alpha S_\nu(\nu_2) \end{equation}
where $L_\nu(\nu_1)$ is the specific luminosity at rest-frame
frequency $\nu_1$ and $S_\nu(\nu_2)$ is the flux density at observed
frequency $\nu_2$.  For the adopted cosmology, chosen to be consistent
with previous work in this field, the luminosity distance
\begin{equation} d_L = (2 c / H_0) (1 + z - \sqrt{1 + z}).
\end{equation} In Fig.~\ref{fig_qsoz4_Lz} we present the results of the
radio surveys, plotting $L_{\rm 1.4~GHz}$ against redshift, with the
radio luminosity cutoff between radio-loud and radio-quiet quasars
indicated by a horizontal dashed line.  Again, consistent with previous
work in this field, we take $\alpha = -0.5$; this is a typical value
for radio-loud quasars.


\subsection{Absolute $B$ Magnitude}

In order to compare our results at $z > 4$ with previous results, we
require transforming the apparent $r$ magnitude, $m_r$ ($\lambda_{\rm
obs} \approx 6400$ \AA), to a restframe absolute $B$ magnitude, $M_B$
($\lambda_{\rm rest} \approx 4400$ \AA).  We follow the methodology of
\markcite{Kennefick:95}Kennefick {et~al.} (1995), in which a model
quasar spectrum is used to relate $m_r$ to the apparent magnitude at
wavelength $1450 (1+z)$ \AA.  A power-law optical spectrum is then
assumed longward of \lya\ to relate this to $M_B$.

The absolute magnitude $M$ is given by \markcite{Hogg:99}(\eg Hogg 1999),
\begin{equation} M_{\rm AB}(\lambda) = m_{\rm AB}(\lambda) - 5 \log
(d_L / 10~ {\rm pc}) + 2.5 \log \left[ (1 + z) { {L_\nu [\lambda / (1 +
z)]} \over {L_\nu (\lambda)}}\right], \end{equation} where both the
apparent magnitude $m$ and the absolute magnitude $M$ are measured at
the same wavelength.  The middle term on the right is called the {\em
distance modulus} and the last term is the {\em k-correction}.
Magnitudes are referred to the $AB$ system \markcite{Oke:74}(Oke 1974), defined as
$m_{\rm AB} (\lambda) \equiv -2.5 \log S_\nu (\lambda) - 48.60$, where
apparent flux density $S_\nu$ is measured in units of \ergscmHz.
Absolute magnitudes at differing wavelengths are related by
\begin{equation} M_{\rm AB}(\lambda_1) - M_{\rm AB}(\lambda_2 ) = -2.5
\log \left[ {{L_\nu(\lambda_1)} \over {L_\nu(\lambda_2 )}} \right] .
\end{equation}  Setting $\lambda_1 = 4400$ \AA\ and $\lambda_2 = 1450
(1+z)$ \AA, the above equations are combined to yield \begin{equation}
M_{\rm AB}(4400) = m_{\rm AB}[1450 (1+z)] - 5 \log (d_L / 10~ {\rm pc})
+ 2.5 \log \left[ (1 + z) { {L_\nu (1450)} \over {L_\nu (4400)}}
\right] .  \end{equation}  We assume a power-law optical spectrum
longward of \lya, $L_\nu \propto \nu^{\rm \alpha_{\rm opt}}$.  For a
standard quasar optical spectral index $\alpha_{\rm opt} = -0.5$
\markcite{Richstone:80, Schneider:92}(\eg Richstone \& Schmidt 1980; Schneider {et~al.} 1992), consistent with previous
work in this field, the offset between Vega-based $M_B$ and $AB$-system
$M_{\rm AB}(4400)$ is given by \markcite{Kennefick:95}Kennefick {et~al.} (1995), \begin{equation}
M_B = M_{\rm AB}(4400) + 0.12.  \end{equation}  We note that the
adopted optical spectral index is likely too flat.  Preliminary results
from near-infrared imaging of $z > 4$ quasars to target the rest-frame
$B$ show substantial scatter in spectral slopes, but with an average
slope slightly steeper than the standard (lower-redshift) value
(Djorgovski \etal, in preparation).  The final expression relating $M_B$
to $m_{\rm AB}[1450(1+z)]$ is then \begin{equation} M_B = m_{\rm
AB}[1450 (1+z)] - 5 \log (d_L / 10~ {\rm pc}) + 2.5 \log (1 + z) + 1.21
\alpha_{\rm opt} + 0.12.  \end{equation}

Our observable is the (Vega-based) $r$ magnitude, $m_r$.  An analytic
expression relating the apparent magnitude at $1450 (1+z)$ \AA\ to
$m_r$ is not possible for the redshift range we are considering as
\lya\ emission and the (redshift-dependent) \lya\ forest traverse the
$r$-band for $4 < z < 5$.  Instead we compute this relation, or
k-correction, \begin{equation} k \equiv m_{\rm AB}[1450(1+z)] - m_{\rm
AB}(r) \end{equation} using the \markcite{Francis:91}Francis {et~al.} (1991) composite (LBQS)
quasar spectrum modified by the \markcite{Madau:95}Madau (1995) model of the hydrogen
opacity of the Universe.  We restrict this calculation to the
\lya\ forest (ignoring the other Lyman series absorptions).  For
$\lambda_{\rm obs} < 1216 (1+z)$ \AA, \markcite{Madau:95}Madau (1995) finds the
optical depth of the intergalactic medium is well-represented as the
sum of \lya-forest absorptions and hydrogen absorption associated with
metal line systems:  respectively, \begin{equation} \tau_{\rm eff} =
0.0036 \left( {\lambda_{\rm obs} \over 1216} \right)^{3.46} + 0.0017
\left( {\lambda_{\rm obs} \over 1216} \right)^{1.68}. \end{equation}
This model well describes the observed flux decrements at the redshifts
considered.


\begin{figure}[!t]\begin{center}
\plotfiddle{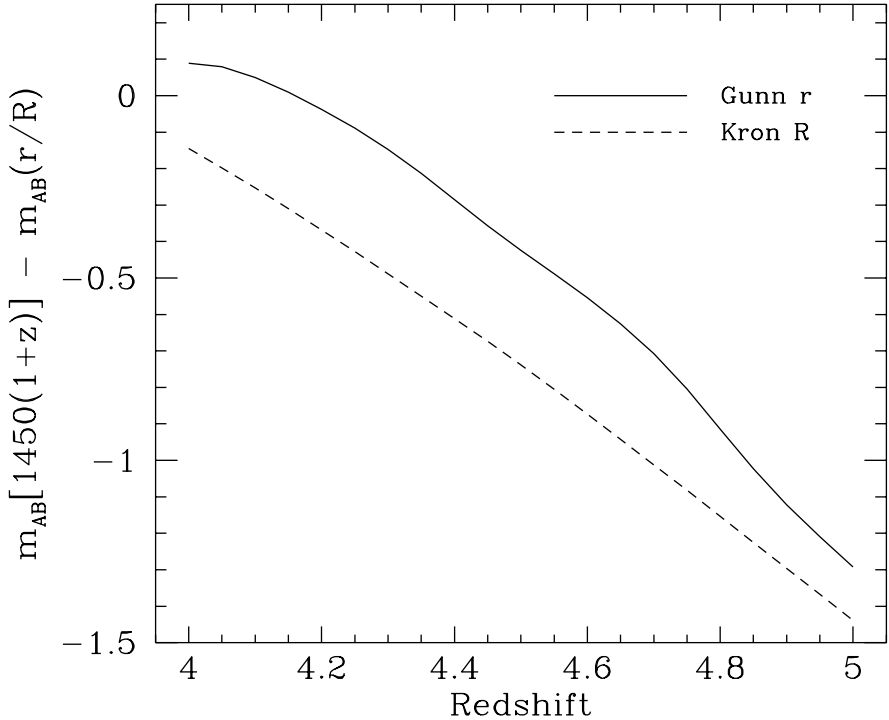}{3.2in}{-90}{50}{50}{-200}{285}
\end{center}

\caption[The k-correction between $1450~(1 + z)$ \AA\ and observed
$r$-band] {The k-correction between apparent $AB$ magnitude at $1450~(1
+ z)$ \AA\ and apparent $AB$ magnitude through Gunn $r$ filter, $m_{\rm
AB}(r)$, plotted as a function of redshift (solid line).  The similar
relation is also plotted for the Kron $R$ filter (dashed line).  As
described in the text, the k-correction was calculated for the
\markcite{Francis:91}Francis {et~al.} (1991) composite LBQS quasar
spectrum modified by the redshift-dependent strength of the
\lya\ forest, as determined from the model of \markcite{Madau:95}Madau
(1995).}

\label{fig_qsoz4_kcor}
\end{figure}

We assume that the \lya\ forest is sparse in the LBQS composite,
relative to the thick jungle we are calculating at $z \geq 4$ and scale
the LBQS composite by $e^{-\tau_{\rm eff}}$ shortward of \lya.  The
k-correction is calculated by convolving the resultant,
redshift-dependent model spectra with the Gunn $r$ filter (see
Fig.~\ref{fig_qsoz4_kcor}).  We show the same curve for the Kron $R$
filter as well; these k-corrections match reasonably well with the
calculations of \markcite{Kennefick:96}Kennefick, Djorgovski, \&  Meylan (1996) for a different model quasar
spectrum and different parameterization of the \lya\ forest.  At the
desired accuracy of these calculations, it is appropriate to adopt the
flat-spectrum (in $S_\nu$) relation between $AB$ and Vega-based $r$
magnitude, $r \equiv m_r \approx m_{\rm AB} (r) - 0.21$, so that our
final expression relating $M_B$ to the observed $m_r$ is
\begin{equation} M_B = m_r - 5 \log (d_L / 10~ {\rm pc}) + 2.5 \log (1
+ z) + 1.21 \alpha_{\rm opt} + k_{\rm eff}, \end{equation} where
$k_{\rm eff} = k + 0.33$.

This method of determining $M_B$ is slightly different from the
approach used by \markcite{Schneider:92}Schneider {et~al.} (1992) in their VLA followup of 22
optically-selected quasars at $z > 3.1$.  \markcite{Schneider:92}Schneider {et~al.} (1992) use the
extinction-corrected spectrophotometric flux density at $1450 \times (1
+ z)$ \AA\ and assume an optical spectral energy index of $-0.5$ to
calculate the absolute $B$ magnitude.  We have recalculated $M_B$ for
their sample using our method.  The root-mean-square (rms)
difference between the methods is 0.30~mag for the 13 quasars above $z
= 4$ in their sample.  If we only consider the nine quasars at $4.0 < z
< 4.4$, the rms difference between the calculated absolute $B$-band
magnitudes is 0.23~mag.  This is comparable to the photometric
uncertainties.  In what follows, we use the revised $M_B$ for the
\markcite{Schneider:92}Schneider {et~al.} (1992) sample.  Absolute $B$ magnitudes should be correct
to $\sim 0.3$ magnitudes.


\begin{figure}[!t]\begin{center}
\plotfiddle{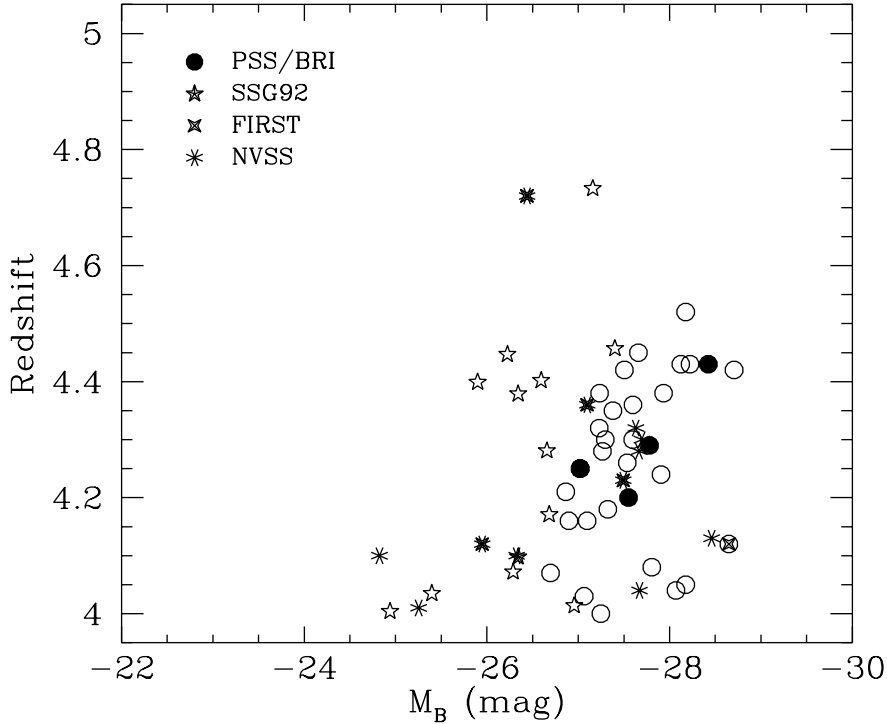}{3.2in}{-90}{50}{50}{-200}{285}
\end{center}

\caption[Location of the $z > 4$ quasars in the $M_B - z$ plane]
{Location of the $z > 4$ quasars in the optical luminosity--redshift
plane.  PSS/BRI refers to the 5~GHz deep imaging discussed in
\S\ref{sec_qsoz4_obs}.  SSG92 refers to the
\markcite{Schneider:92}Schneider {et ~al.} (1992) sample.
Non-detections are indicated with open symbols.  The $z > 4$ FIRST and
NVSS non-detections are not presented.}

\label{fig_qsoz4_zMB1}
\end{figure}

\subsection{Coverage of the Optical Luminosity--Redshift Plane}

\begin{figure}[!t]\begin{center}
\plotfiddle{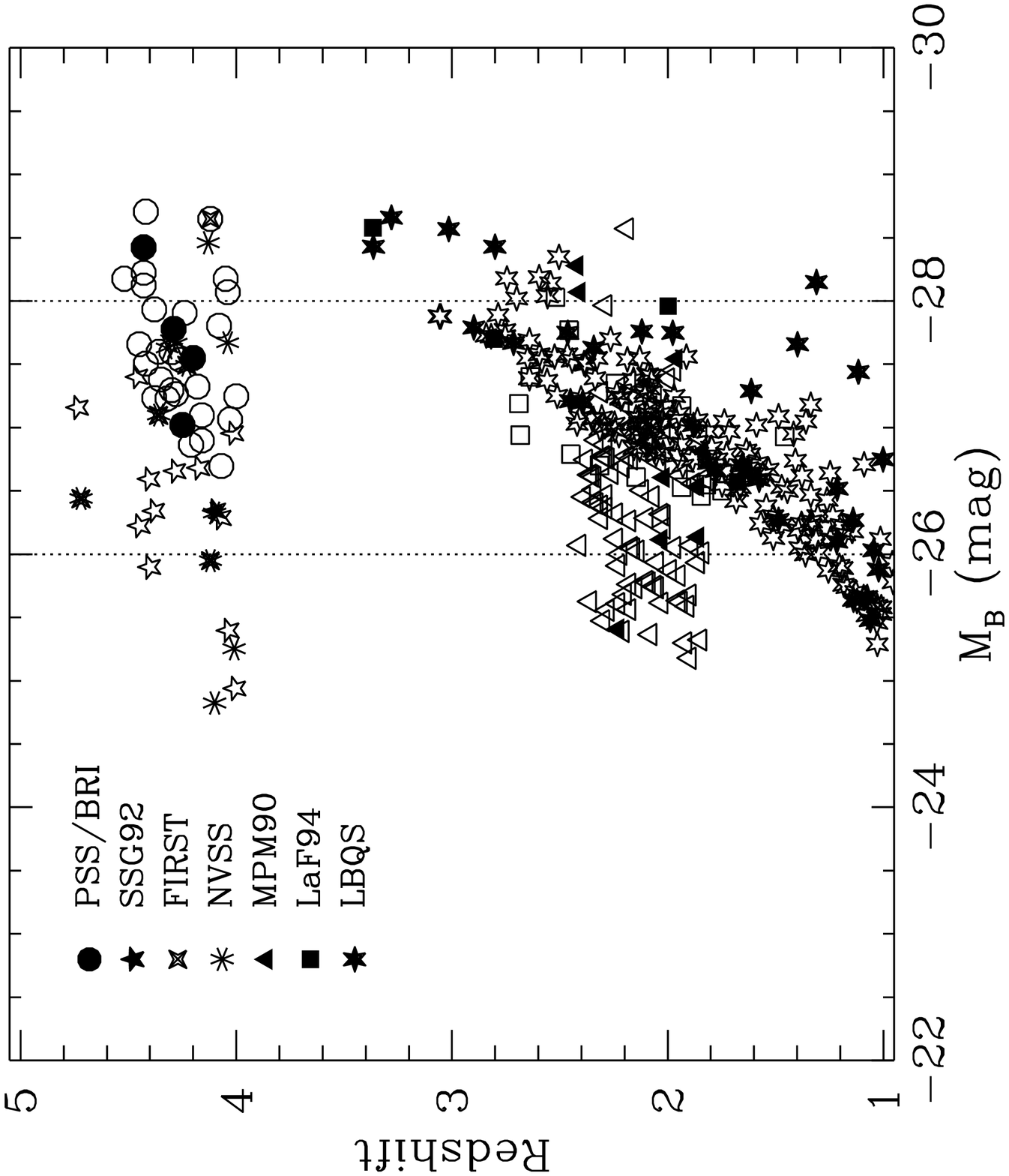}{3.2in}{-90}{50}{50}{-200}{285}
\end{center}

\caption[Location of optically-selected quasars in $M_B - z$ plane]
{Location of all quasars considered here in the optical
luminosity--redshift plane.  PSS/BRI refers to the 5~GHz deep imaging
discussed in \S\ref{sec_qsoz4_obs}.  SSG92 refers to the
\markcite{Schneider:92}Schneider {et~al.} (1992) sample.  MPM90 refers
to \markcite{Miller:90}Miller {et~al.} (1990).  LaF94 refers to
\markcite{LaFranca:94}{La~Franca} {et~al.} (1994).  LBQS refers to the
 8.4~GHz VLA imaging of LBQS quasars reported in
\markcite{Visnovsky:92}Visnovsky {et~al.} (1992) and
\markcite{Hooper:95}Hooper {et~al.} (1995).  Non-detections are
indicated with open symbols.  The $z > 4$ FIRST and NVSS non-detections
are not presented.  Vertical dotted lines indicate the optical
luminosity range considered in our analysis.  }

\label{fig_qsoz4_zMB2}
\end{figure}

Fig.~\ref{fig_qsoz4_zMB1} summarizes the location of the $z > 4$
quasars in the optical luminosity--redshift plane.  Consonant with
their extreme distances, these quasars have extremely bright absolute
magnitudes, $M_B \simlt -26$.  To probe evolution in the radio-loud
quasar fraction, we require comparison with a sample of comparably
luminous quasars at lower redshift.  Fig.~\ref{fig_qsoz4_zMB2}
illustrates such a sample:  we have augmented the $z > 4$ quasars with
several lower-redshift surveys from the literature, summarized below.
As before, k-corrections have been calculated assuming that both the
radio spectral index $\alpha$ and the optical spectral index
$\alpha_{\rm opt}$ are equal to $-0.5$.  We do not include radio
surveys of optically-selected quasars which have few objects at our
target optical luminosity \markcite{Goldschmidt:99}(\eg\ the Edinburgh quasar sample
considered in Goldschmidt {et~al.} 1999).


{\em Miller sample:}  \markcite{Miller:90}Miller {et~al.} (1990) report on sensitive 5~GHz VLA
observations of 105 quasars at $1.8 < z < 2.5$.  They provide apparent
magnitudes at the restframe wavelength of 1475 \AA.  Following
\markcite{LaFranca:94}{La~Franca} {et~al.} (1994) and \markcite{Goldschmidt:99}Goldschmidt {et~al.} (1999), we transform to $B$
using $m_B = m_{1475} + 0.23$; this is correct at the average redshift
of the sample for the assumed optical spectral index.  The typical
limiting flux density of this survey is $S_{\rm 1.4 GHz} = 0.5$ mJy ($3
\sigma$).  For $z \simeq 2$, this translates to a radio luminosity of
$L_{\rm 1.4 GHz} \simeq 1.5 \times 10^{32}~ h_{50}^{-2}~$ \ergsHz, well
below the radio-loud cutoff.

{\em La~Franca sample:} \markcite{LaFranca:94}{La~Franca} {et~al.} (1994) report on 5~GHz VLA
observations of 23 optically-selected quasars with $B < 19.4$ at
intermediate redshift ($0.8 < z < 3.4$).  The average $3 \sigma$  limits
to detections are $\sim 0.12$ mJy, corresponding to $L_{\rm 1.4 GHz}
\simeq 8 \times 10^{31}~ h_{50}^{-2}~$ \ergsHz at $z \simeq 3$.  This
is well below the radio-loud cutoff.  We use the absolute luminosities
provided in the paper.

{\em LBQS sample:}  256 quasars from the LBQS have been observed at
8.4~GHz by the VLA by \markcite{Visnovsky:92}Visnovsky {et~al.} (1992) and \markcite{Hooper:95}Hooper {et~al.} (1995).  This
large survey, sampling $0.2 \simlt z \simlt 3.5$ and $-22.5 \simgt M_B
\simgt -29$, suffers from strongly correlated luminosity and redshift,
as expected for a flux-limited sample:  at a given redshift, $M_B$ only
spans $\approx 1.5$ magnitudes.  We use the absolute magnitudes
provided in the papers.  The median $3 \sigma$ noise limit of $S_{\rm
8.4 GHz} = 0.29$ mJy corresponds to $L_{\rm 1.4 GHz} \simeq 2.5 \times
10^{32}~ h_{50}^{-2}~$ \ergsHz, again below the radio-loud cutoff.
We do not include the additional 103 LBQS quasars with 8.4~GHz VLA
observations discussed in \markcite{Hooper:96}Hooper {et~al.} (1996) as that sample is largely
optically less-luminous, lower-redshift quasars.


\begin{figure}[!t]\begin{center}
\plotfiddle{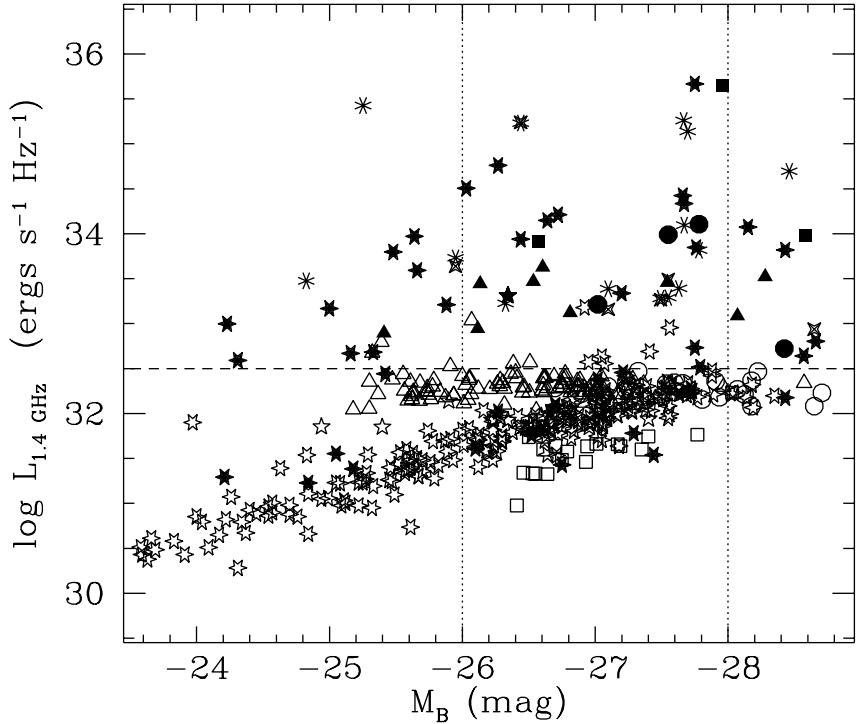}{3.2in}{-90}{50}{50}{-200}{285}
\end{center}

\caption[Location of optically-selected quasars in $M_B - L_{\rm 1.4
GHz}$ plane] {Location of all quasars considered here in the optical
luminosity--radio luminosity plane.  Symbols are as in
Fig.~\ref{fig_qsoz4_zMB2}; again, non-detections are indicated with
open symbols.  Horizontal dashed line refers to the cutoff between
radio-loud and radio-quiet quasars.  A few sources with elevated noise
levels are classified as radio-loud.  Vertical dotted lines indicate
the optical luminosity range considered in our analysis.  }

\label{fig_qsoz4_MBL1}
\end{figure}

Fig.~\ref{fig_qsoz4_MBL1} illustrates the location of the quasars in
$M_B - L_{\rm 1.4 GHz}$ space.  For the purposes of the following
analysis, we assume non-detections have radio fluxes equal to their $3
\sigma$ noise value.  We omit the new $z > 4$ radio-loud quasars
detected by FIRST and NVSS in the following analysis, as those surveys
are insufficiently sensitive to reach the radio-loud cutoff at $z >
4$.  The detection limits of the other surveys are sufficiently deep
that few non-detections are above the radio-loud/radio-quiet boundary;
we conservatively classify those few sources as radio-loud below.  Our total
sample is 428 quasars spanning $0.2 < z < 4.7$, $-22.7 > M_B > -28.7$,
and $30.08 < \log L_{\rm 1.4 GHz} (h_{50}^{-2} \ergsHz) < 35.7$.


\begin{figure}[!t]\begin{center}
\plotfiddle{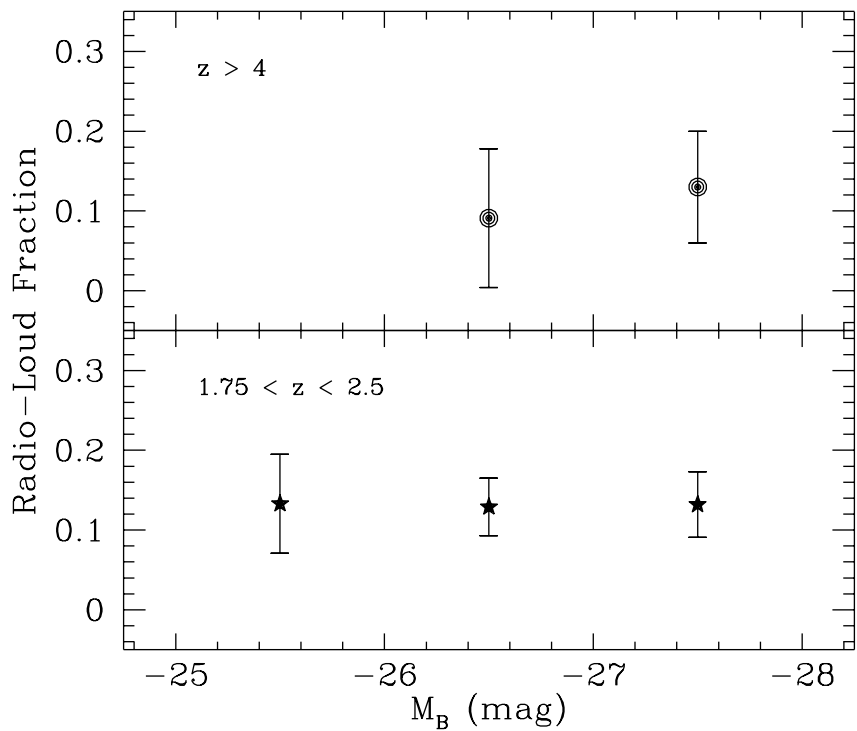}{3.2in}{-90}{50}{50}{-200}{285}
\end{center}

\caption[Radio-loud fraction as function of absolute magnitude]
{Fraction of radio-loud quasars as function of absolute magnitude.  Two
redshift ranges are plotted with 1~mag wide bins.  No evidence for
luminosity dependence in the radio-loud fraction is evident from this
limited data set.}

\label{fig_qsoz4_FrlqMB}
\end{figure}

\section{Radio-Loud Fraction}
\label{sec_qsoz4_discuss}

\subsection{Radio-Loud Fraction as a Function of Optical Luminosity}

We first consider if we detect optical luminosity dependence in the
radio-loud fraction.  We consider two redshift ranges where we have
large samples of quasars, and divide the sample into absolute magnitude
bins where the quasars are {\em approximately} smoothly distributed.
Fig.~\ref{fig_qsoz4_FrlqMB} shows this fraction for each redshift bin.
Error bars shown are the square root of the variance, $f (1-f) / N$,
where $f$ is the radio-loud fraction and $N$ is the number of quasars
in the bin considered \markcite{Schneider:92}(\eg Schneider {et~al.} 1992).  The 
impression from Fig.~\ref{fig_qsoz4_FrlqMB} is that for a given
redshift bin, the radio-loud fraction is independent of optical
luminosity.  This result stands in contrast to the analysis of
\markcite{Goldschmidt:99}Goldschmidt {et~al.} (1999) who find that the radio-loud fraction increases
with luminosity for each redshift bin considered.  Consideration of the
radio-loud fraction plotted in Fig.~7 of their paper suggests that the
optical luminosity dependence claimed at $1.3 < z < 2.5$ depends
largely on the poorly measured radio-loud fraction at $M_B = -28$.
However, at $0.3 < z < 1.3$, their data convincingly shows optical
luminosity dependence in the radio-loud fraction.  Comparing the
ordinate between the two panels of Fig.~\ref{fig_qsoz4_FrlqMB} suggests
that the radio-loud fraction remains approximately constant with redshift; we
consider this next.


\begin{figure}[!t]\begin{center}
\plotfiddle{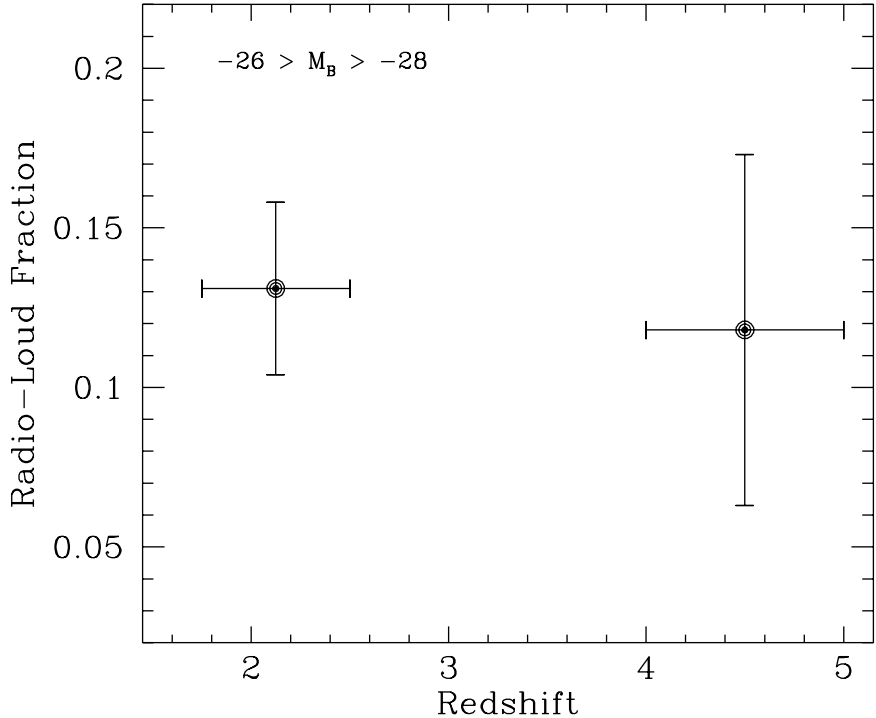}{3.2in}{-90}{50}{50}{-200}{285}
\end{center}

\caption[Radio-loud fraction as function of redshift] {Fraction of
radio-loud quasars as function of redshift.  Two redshift ranges are
plotted.  No redshift dependence in the radio-loud fraction is
evident.}

\label{fig_qsoz4_Frlqz}
\end{figure}

\subsection{Radio-Loud Fraction as a Function of Redshift}

Since there is little evidence of optical luminosity dependence of the
radio-loud fraction in our data set, we decrease our errors by
considering the radio-loud fraction in a larger absolute magnitude
range, $-26 < M_B < -28$.  Fig.~\ref{fig_qsoz4_Frlqz} shows the results
of this analysis, with error bars calculated as above.  At $1.75 < z <
2.5$, we classify 20 out of the 153 quasars in the optical luminosity
range considered as radio-loud, corresponding to $13.1 \pm 2.7$\%\ of
the quasars.  At $z > 4$, we classify 4 out of the 34
optically-selected quasars as radio-loud (considering only the
\markcite{Schneider:92}Schneider {et~al.} (1992) and our targeted VLA survey).  This corresponds to
$11.8 \pm 5.5$\%\ of the quasars being radio-loud.  No evolution in the
radio-loud fraction is detected.

The FIRST/NVSS detections of optically-selected quasars at $z > 4$ also
provides a lower limit to the radio-loud fraction at early cosmic
epoch.  Of the 107 $z > 4$ optically-selected quasars, 79 have optical
luminosities $-26 > M_B > -28$.  From this restricted sample, 35
overlap with FIRST of which one was detected, implying a statistically
unrobust radio-loud fraction $> 3$\%.  All 79 quasars in the
luminosity range considered overlap with NVSS; four were detected,
implying a radio-loud fraction $> 5$\% at $z > 4$.

\section{Conclusions}

We report on two programs to study the radio-properties of
optically-selected quasars at high-redshift.  First, we consider deep,
targeted 5~GHz imaging of 32 $z > 4$ quasars selected from the Palomar
multicolor quasar survey.  Four sources are detected.  We also
correlate a comprehensive list of 134 $z > 4$ quasars, entailing all
such sources we are aware of as of mid-1999, with two deep 1.4~GHz VLA
sky surveys.  We find five new radio-loud quasars, not including one
quasar identified in the targeted program.  In total, we report on
eight new radio-loud quasars at $z > 4$; only seven such sources are in
the literature currently.

We use this new census to probe the evolution of the radio-loud
fraction with redshift.  We find that, for $-25 \simgt M_B \simgt -28$
and $2 \simlt z \simlt 5$, radio-loud fraction is independent of
optical luminosity.  We also find no evidence for radio-loud fraction
depending on redshift.  If the conventional wisdom that radio-loud AGN
are preferentially identified with early-type galaxies remains robust
at high redshift, this result could have implications regarding the
formation epoch of late-type versus early-type galaxies.  In
hierarchical models of galaxy formation, one expects the late-type
(less massive) systems to form first.  Mergers are required to form the
early-type (more massive) systems.  Eventually, at high enough
redshift, one would then expect the radio-loud fraction of AGN to fall
precipitously.  Our results show this epoch lies beyond $z \simeq 4$,
providing further evidence for an early formation epoch for early-type
galaxies.

\acknowledgements

We acknowledge the efforts of the POSS-II and DPOSS teams, which
produced the PSS sample used in this work, and, in particular, N. Weir,
J. Kennefick, R. Gal, S. Odewahn, R. Brunner, V. Desai, and J.
Darling.  The DPOSS project is supported by a generous grant from the
Norris Foundation.  The VLA of the National Radio Astronomy Observatory
is operated by Associated Universities, Inc., under a cooperative
agreement with the National Science Foundation.  We thank Hyron Spinrad
and Pippa Goldschmidt for useful comments and Chris Fassnacht for
interesting discussion.  DS acknowledges support from IGPP/Livermore
grant IGPP/LLNL 98-AP017.  SGD acknowledges partial support from the
Bressler Foundation.


\end{document}